%% file: main.tex
\documentclass[sigconf]{acmart}
\copyrightyear{2025}
\acmYear{2025}
\setcopyright{cc}
\setcctype{by}
\acmConference[WWW Companion '25]{Companion Proceedings of the ACM Web Conference 2025}{April 28-May 2, 2025}{Sydney, NSW, Australia}
\acmBooktitle{Companion Proceedings of the ACM Web Conference 2025 (WWW Companion '25), April 28-May 2, 2025, Sydney, NSW, Australia}
\acmDOI{10.1145/3701716.3715516}
\acmISBN{979-8-4007-1331-6/25/04}

\makeatletter
\gdef\@copyrightpermission{
  \begin{minipage}{0.3\columnwidth}
   \href{https://creativecommons.org/licenses/by/4.0/}{\includegraphics[width=0.90\textwidth]{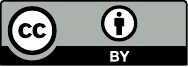}}
  \end{minipage}\hfill
  \begin{minipage}{0.7\columnwidth}
   \href{https://creativecommons.org/licenses/by/4.0/}{This work is licensed under a Creative Commons Attribution International 4.0 License.}
  \end{minipage}
  \vspace{5pt}
}
\makeatother

\usepackage{cleveref}
\crefname{section}{§}{§§} 
\usepackage{enumitem}
\usepackage{multirow}
\usepackage{colortbl}
\usepackage[normalem]{ulem}
\useunder{\uline}{\ul}{}
\usepackage{adjustbox}
\usepackage{xspace}
\usepackage{subcaption}

\newcommand{\proposed}{CDiff4Rec\xspace}
\newcommand{\smallsection}[1]{\noindent\textbf{#1}}

\title{Collaborative Diffusion Model for Recommender System}

\author{Gyuseok Lee}
\affiliation{
    \institution{Pohang University of \\ Science and Technology}
    \city{Pohang}
    \state{Gyeongbuk\\}
    \country{Republic of Korea}
}
\email{gyuseok.lee@postech.ac.kr}

\author{Yaochen Zhu}
\affiliation{
    \institution{University of Virginia}
    \city{Charlottesville}
    \state{VA}
    \country{USA}
}
\email{uqp4qh@virginia.edu}

\author{Hwanjo Yu}
\affiliation{
    \institution{Pohang University of \\ Science and Technology}
    \city{Pohang}
    \state{Gyeongbuk\\}
    \country{Republic of Korea}
}
\email{hwanjoyu@postech.ac.kr}
\authornote{Corresponding author.}

\author{Yao Zhou}
\affiliation{
    \city{}
    \institution{Google Inc.}
    \city{Mountain View}
    \state{CA}
    \country{USA}
}
\email{yaozhoucosmos@google.com}

\author{Jundong Li}
\affiliation{
    \institution{University of Virginia}
    \city{Charlottesville}
    \state{VA}
    \country{USA}
}
\email{jundong@virginia.edu}

\ccsdesc[500]{Information systems~Recommender systems}

\keywords{Diffusion Model, Collaborative Filtering, Item-Side Information}

\begin{document}

\begin{abstract}
\input{Section/Abstract}
\end{abstract}

\maketitle

\section{Introduction}
\input{Section/Introduction}

\vspace{-0.2cm}
\section{Preliminary}
\input{Section/Preliminary_onlydiff}


\begin{figure*}[h]
    \centering
    \includegraphics[width=0.91\textwidth]{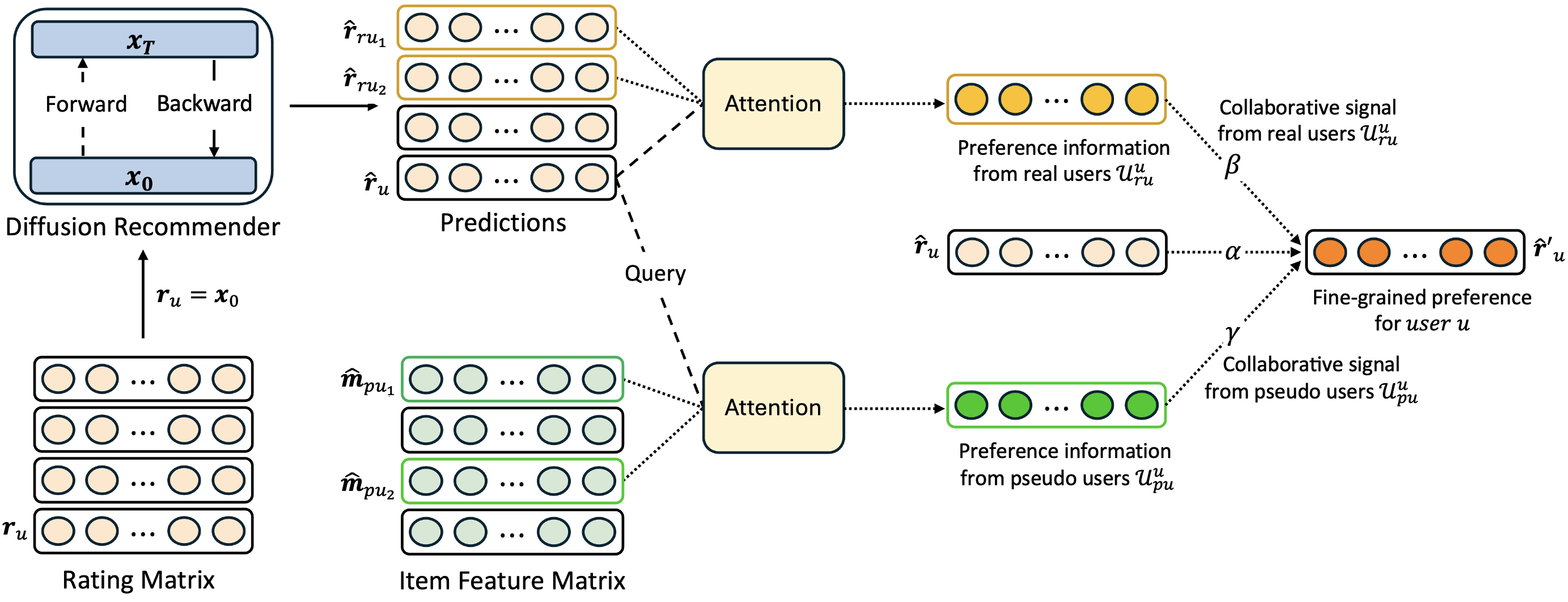}
    \caption{The overview of the proposed Collaborative Diffusion Model for Recommender System (\proposed).}
    \label{fig:main_figure}
\vspace{-0.4cm}
\end{figure*}

\vspace{-0.2cm}
\section{\proposed}
\input{Section/Method}
\input{Table/dataset}

\section{Experiments}

\input{Section/Experiment}

\section{Conclusion}

\input{Section/Conclusion}

\section*{Acknowledgement}

Gyuseok Lee and Hwanjo Yu were supported by the IITP grant funded by the MSIT (South Korea, No.2018-0-00584, RS-2019-II191906), the NRF grant funded by the MSIT (South Korea, No.RS-2023-00217286, No.RS-2024-00335873), as the other authors were not funded by these grants.

\bibliographystyle{ACM-Reference-Format}
\balance
\bibliography{main.bib}



\end{document}

%% file: Section/Abstract.tex
Diffusion-based recommender systems (DR) have gained increasing attention for their advanced generative and denoising capabilities. However, existing DR face two central limitations: \textbf{\textit{(i)}} a trade-off between enhancing generative capacity via noise
injection and retaining the loss of personalized information.
\textbf{\textit{(ii)}} the underutilization of rich item-side information.
To address these challenges, we present a \textbf{\underline{C}}ollaborative \textbf{\underline{Diff}}usion model for \textbf{\underline{Rec}}ommender System (\textbf{\proposed}). 
Specifically, \proposed generates \textit{pseudo-users} from item features and leverages collaborative signals from both real and pseudo personalized neighbors identified through behavioral similarity, thereby effectively reconstructing nuanced user preferences.
Experimental results on three public datasets show that \proposed outperforms competitors by effectively mitigating the loss of personalized information through the integration of item content and collaborative signals. 

%% file: Section/Introduction.tex
Recommender systems have become indispensable tools for 
helping users make informed decisions by delivering personalized recommendations amidst the overwhelming information on the Web~\cite{ricci2010introduction,zhu2024collaborative}. A widely adopted strategy in RS is collaborative filtering (CF), which models user preferences based on user-item interactions through traditional methods (e.g., matrix factorization~\cite{koren2009matrix, he2017neural} and graph neural networks~\cite{wang2019neural, he2020lightgcn}). 
Moreover, deep generative models further enhance CF by effectively capturing user preferences in noisy and sparse interactions through their generative capabilities~\cite{harshvardhan2020comprehensive,wu2016collaborative} (e.g., Generative Adversarial Networks (GANs)~\cite{wang2017irgan, chae2018cfgan} and Variational Autoencoders (VAEs)~\cite{liang2018variational,zhu2022mutually}).
However, these generative models face significant challenges: GANs are prone to unstable training~\cite{xu2020understanding}, while VAEs often suffer from posterior collapse, limiting their ability to model user preferences accurately~\cite{liu2022hybrid}. 

Recently, diffusion models introduce a novel generation paradigm
by progressively adding Gaussian noise to inputs \textit{(forward process)} and reconstructing them via a parameterized Markov chain \textit{(backward process)}.
With their capability to significantly enhance generative quality and stability~\cite{ho2020denoising, luo2022understanding}, 
diffusion-based recommender systems (DR) have been explored for higher-quality recommendations~\cite{yang2024generate, li2023diffurec, du2023sequential, wu2023diff4rec, liu2023diffusion, wang2023diffusion, han2024controlling}. 
For example, DreamRec~\cite{yang2024generate} utilizes guided diffusion to generate oracle items in sequential recommendation (SR). DiffuASR~\cite{liu2023diffusion} mitigates the challenges of data sparsity and long-tail user problems in SR.
DiffRec~\cite{wang2023diffusion} is the first to apply DR to CF by denoising corrupted user-item interactions.

However, existing DRs still encounter two major challenges: 
\textbf{\textit{(i)}} a dilemma between maximizing generative capacity via noise injection and the loss of personalized information, and \textbf{\textit{(ii)}} the underutilization of item-side information.
First, DR faces a dilemma: Excessive noise-adding operations transform inputs into pure noise, causing substantial loss of personalized information and complicating the reconstruction of user preferences, whereas insufficient noise limits generative capabilities.
Although reducing noise levels or diffusion steps can partially reconcile this trade-off~\cite{wang2023diffusion}, performance remains highly sensitive to hyperparameters and prone to inconsistency.
Second, the user-oriented nature of DR leaves the integration of rich item-side information (e.g., item reviews) underexplored, limiting their potential to fully leverage complementary knowledge sources for more accurate recommendations.

To address these challenges, we propose a \textbf{\underline{C}}ollaborative \textbf{\underline{Diff}}usion model for \textbf{\underline{Rec}}ommender System (\textbf{\proposed}). 
The goal of \proposed is to mitigate the loss of personalized information by leveraging item-side information and collaborative signals (i.e., latent knowledge derived from behavioral similarity between users).
Specifically, \proposed operates in three main steps: 
First, we generate \textit{pseudo-users} from item features (i.e., review words) by treating each feature (i.e., a word) as a pseudo-user based on its co-occurrence across multiple items (e.g., a word appears in the reviews of different items).
Thus, the interaction history of a pseudo-user is analogous to a user's interaction history spanning multiple items.
Second, for each real user (query user), we identify two types of personalized top-$K$ neighbors—one from real users and another from pseudo-users—based on behavioral similarity.
Finally, during the diffusion process, we incorporate the preference information from both real and pseudo neighbors to 
generate a fine-grained representation of the query user's preferences.
This integration of item content and collaborative signals substantially enhances the cascade reconstruction in DR.
Experimental results on three public datasets show that \proposed outperforms various baselines.
Our additional analyses support the effectiveness of each proposed component.

%% file: Section/Preliminary_onlydiff.tex
\smallsection{Notations.}
We focus on collaborative filtering with implicit feedback \cite{hu2008collaborative}. Let $\mathcal{R} \in \{0, 1\}^{|\mathcal{U}| \times |\mathcal{I}|}$ be a binary rating matrix, where $\mathcal{U}$ and $\mathcal{I}$ are sets of users and items, respectively. The $u$-th row of $\mathcal{R}$, denoted as $\mathbf{r}_u \in \{0, 1\}^{|\mathcal{I}|}$, is the interaction history of user $u$. 
Let $\mathcal{M} \in \mathbb{R}^{|\mathcal{F}| \times |\mathcal{I}|}$ be an item feature matrix, where $\mathcal{F}$ is the set of distinct features.
In this paper, we treat $\mathcal{F}$ as words extracted from item reviews.
Each element $m_{fi}$ is the count of word $f$ for item $i$.

\smallsection{Diffusion-based recommender systems (DR)~\cite{wang2023diffusion}.}
We outline the forward, backward processes, training, and inference of DR:
\begin{itemize}[leftmargin=*]
    \item \textbf{Forward process.}
    During the forward process, the interaction history of user $u$ (i.e., $\mathbf{r}_u$) is treated as the input $\mathbf{x}_0$ at timestep $t = 0$. To make the reconstruction more robust, Gaussian noise is progressively added to $\mathbf{x}_0$ over $T$ timesteps, resulting in $\mathbf{x}_t$:
    \vspace{-0.2cm}
    \begin{equation}
        \begin{aligned}
            q\left(\mathbf{x}_t \mid \mathbf{x}_0\right)=\mathcal{N}\left(\mathbf{x}_t ; \sqrt{\bar{\alpha}_t} \mathbf{x}_0,\left(1-\bar{\alpha}_t\right) \boldsymbol{I}\right),
            \label{forward}
        \end{aligned}
    \vspace{-0.2cm}
    \end{equation}
    where $\bar{\alpha}_t = \prod_{t'=1}^{T} \alpha_{t'}$, and $\alpha_{t'}$ is a hyperparameter that controls the level of Gaussian noise added at timestep $t'$.

    \item \textbf{Backward process.}
    The backward process reconstructs the input step-by-step using a parameterized Markov chain $\theta$ that approximates the tractable forward process posteriors $q\left(\mathbf{x}_{t-1} \mid \mathbf{x}_t, \mathbf{x}_0\right)$:
    \vspace{-0.1cm}
    \begin{equation}
        \begin{aligned}
            p_\theta(\mathbf{x}_{t-1}|\mathbf{x}_t) &= \mathcal{N} (\boldsymbol{\mu}_\theta(\mathbf{x}_t, t), 
            \sigma^2(t)\mathbf{I}), \text{ where}\\
            \boldsymbol{\mu}_\theta\left(\mathbf{x}_t, t\right)&=\frac{\sqrt{\alpha_t}\left(1-\bar{\alpha}_{t-1}\right)}{1-\bar{\alpha}_t} \mathbf{x}_t+\frac{\sqrt{\bar{\alpha}_{t-1}}\left(1-\alpha_t\right)}{1-\bar{\alpha}_t} \hat{\mathbf{x}}_\theta\left(\mathbf{x}_t, t\right).
        \end{aligned}
    \end{equation}
    
    \item \textbf{Training.}
    Basically, diffusion models~\cite{ho2020denoising} is trained by maximizing the Evidence Lower Bound of the log-likelihood of $\mathbf{x}_0$:
    \small{
        \begin{align}
        \log p(\mathbf{x}_0) & \geq \underbrace{\mathbb{E}_{q\left(\mathbf{x}_1 \mid \mathbf{x}_0\right)}\left[\log p_{\theta}\left(\mathbf{x}_0 \mid \mathbf{x}_1\right)\right]}_{\text {Reconstruction term }}-\underbrace{D_{\mathrm{KL}}\left(q\left(\mathbf{x}_T \mid \mathbf{x}_0\right) \| p\left(\mathbf{x}_T\right)\right)}_{\text {Prior matching term}} \nonumber \\
        & -\sum_{t=2}^T \underbrace{\mathbb{E}_{q\left(\mathbf{x}_t \mid \mathbf{x}_0\right)}\left[D_{\mathrm{KL}}\left(q\left(\mathbf{x}_{t-1} \mid \mathbf{x}_t, \mathbf{x}_0\right) \| p_\theta\left(\mathbf{x}_{t-1} \mid \mathbf{x}_t\right)\right)\right].}_{\text {Denoising matching term}}
    \end{align}
    }
    \normalsize
    Meanwhile, DR is simply optimized using the following loss:
    \begin{equation}\label{eq:diffrec_loss}
    \begin{aligned}
        \mathcal{L}_t = &\,\mathbb{E}_{\boldsymbol{q}\left(\textbf{x}_t \mid \textbf{x}_0\right)}\left[C\left\|\hat{\textbf{x}}_\theta\left(\textbf{x}_t, t\right)-\textbf{x}_0\right\|_2^2\right],
    \end{aligned}\vspace{-0.1cm}
    \end{equation}

    where $C = \frac{1}{2}\left(\frac{\bar{\alpha}_{t-1}}{1-\bar{\alpha}_{t-1}}-\frac{\bar{\alpha}_t}{1-\bar{\alpha}_t}\right)$, and  $\hat{\textbf{x}}_\theta\left(\textbf{x}_t, t\right)$ is the only component in DR, implemented as a Multi-Layer Perceptron (MLP).
    
    \vspace{0.15cm}
    \item \textbf{Inference.} After training $\theta$, the corrupted input $\mathbf{x}_t$ is iteratively denoised $\mathbf{x}_t \rightarrow \mathbf{x}_{t-1} \rightarrow \cdots \rightarrow \mathbf{x}_0$ by leveraging $p_\theta(\mathbf{x}_{t-1}|\mathbf{x}_t)$.

\end{itemize}

%% file: Section/Method.tex
In this section, we introduce a \textbf{\underline{C}}ollaborative \textbf{\underline{Diff}}usion model for \textbf{\underline{Rec}}ommender System (\textbf{\proposed}) (see Figure~\ref{fig:main_figure}). 
CDiff4Rec consists of three stages: \textbf{\textit{(i)}} generation of pseudo-users from item review data (\cref{sub:pseudo_user}), \textbf{\textit{(ii)}} identification of two types of personalized top-$K$ neighbors from real and pseudo-users (\cref{sub:ident_topk}), and \textbf{\textit{(iii)}} aggregation of collaborative signals from those neighbors by incorporating their preference information during the diffusion process (\cref{sub:agg_col}).

\vspace{-0.2cm}
\subsection{Pseudo-User Generation from Item Reviews}
\label{sub:pseudo_user}
To effectively mitigate the loss of personalized information in user-oriented DR, 
we strategically leverage item features.
Specifically, we generate pseudo-users by treating each feature (i.e., a review word) as an individual user with a unique interaction history.
Let $pu$ denote a pseudo-user corresponding to a specific feature $f$ (i.e., $pu = f \in \mathcal{F}$), and $ru$ denote a real user (i.e., $ru = u \in \mathcal{U}$).
For each pseudo-user, we construct a vector $\mathbf{m}_{pu} \in [0, 1]^{|I|}$ by processing the original feature vector $\mathbf{m}_f \in \mathbb{R}^{|\mathcal{I}|}$ as follows: First, we apply TF-IDF to compute the importance of the feature for each item.
Then, we perform Min-Max normalization to scale values into $[0, 1]$.

Here, $\mathbf{m}_{pu}$ is the normalized interaction vector of a pseudo-user, which encodes the relevance of feature $f$ across all items. 
This is analogous to a real user’s binary interaction vector $\mathbf{r}_{ru} \in \{0, 1\}^{|\mathcal{I}|}$, which is the co-visit history of a user $u$ across all items.
We emphasize two advantages of leveraging item features as pseudo-users: 
First, $\mathbf{m}_{pu}$ can be seamlessly incorporated into the existing CF framework based on implicit feedback without requiring specialized loss functions for item features.
Second, introducing pseudo-users into DR yields mutual benefits:
DR’s advanced denoising capabilities can reduce noise in the normalized pseudo-user vectors, selectively leveraging the most salient item content.
Meanwhile, the rich item-side information from pseudo-users complements user-oriented signals, contributing to a more accurate reconstruction of user preferences.
Thus, using DR with pseudo-users yields synergistic effects, ultimately enhancing recommendation quality.



\vspace{-0.2cm}
\subsection{Personalized Top-$K$ Neighbors Identification}
\label{sub:ident_topk}
With the newly generated pseudo-users, we identify two personalized user sets for each query user $u \in \mathcal{U}$: one from real users and the other from pseudo-users.
To determine the top-$K$ neighbors based on behavior similarity, we employ a distance function $\phi: \mathbb{R}^{|\mathcal{I}|} \times \mathbb{R}^{|\mathcal{I}|} \rightarrow \mathbb{R}$. 
In this work, we use cosine distance as: 
\vspace{-0.1cm}
\begin{equation}
    \begin{aligned}
        \mathcal{U}^u_{ru} & = \{ru \,|\, \underset{ru \in \mathcal{U}}{\text{argsort }} \phi(\mathbf{r}_u, \mathbf{r}_{ru})[:K] \}, \\
        \mathcal{U}^u_{pu} & = \{pu \,|\, \underset{pu \in \mathcal{F}}{\text{argsort }} \phi(\mathbf{r}_u, \mathbf{m}_{pu})[:K] \},
    \end{aligned}
    \vspace{-0.1cm}
    \label{make_user}
\end{equation}
where $\mathcal{U}^u_{ru}$ and $\mathcal{U}^u_{pu}$ are the top-$K$ neighbors from real and pseudo-users for query user $u$.
Note that $\mathcal{U}^u_{ru}$ and $\mathcal{U}^u_{pu}$ 
are precomputed and cached during preprocessing, allowing for efficient utilization without additional computation in model training and inference.
\vspace{-0.2cm}
\subsection{Collaborative Signals Aggregation Using PI}
\label{sub:agg_col}
Throughout the diffusion process (forward and then backward), DR generates user interactions by
transforming $\mathbf{r}_u$ into $\hat{\mathbf{r}}_u$, which is represented as $\hat{\textbf{x}}_\theta\left(\textbf{x}_t, t\right) = \hat{\mathbf{r}}_u$ (Eq.~\eqref{eq:diffrec_loss}). 
However, $\hat{\mathbf{r}}_u$ is insufficient because excessive noise erodes personalized information, while limited noise reduces the generative capabilities required to capture latent user preferences.
To address this, we aggregate collaborative signals from two sets of top-$K$ neighbors (i.e., $\mathcal{U}^u_{ru}$ and $\mathcal{U}^u_{pu}$) by incorporating their preference information (PI) during the diffusion process.
This approach yields a fine-grained representation of user preferences, denoted as ${\hat{\mathbf{r}}_u}'$, as follows:
\vspace{-0.1cm}
\begin{equation}
\label{eq:fg_obj}
    \begin{aligned}
        {\hat{\mathbf{r}}_u}' & = \alpha \hat{\mathbf{r}}_u + 
        \beta \underbrace{\sum_{ru_i \in \mathcal{U}^u_{ru}} a_{ru_i}\hat{\mathbf{r}}_{ru_i}}_{\text {PI from real neighbors}} +
        \gamma \underbrace{\sum_{pu_j \in \mathcal{U}^u_{pu}} a_{pu_j}\hat{\mathbf{m}}_{pu_j}}_{\text {PI from pseudo neighbors}},
    \end{aligned}\vspace{-0.1cm}
\end{equation}
where $\alpha$, $\beta$, and $\gamma$ (with $\alpha+\beta+\gamma = 1$) are hyperparameters that control the weights of the user's own, real neighbors, and pseudo neighbors. 
$\hat{\mathbf{r}}_{ru_i} \in \mathbb{R}^{|\mathcal{I}|}$ and $\hat{\mathbf{m}}_{pu_j} \in \mathbb{R}^{|\mathcal{I}|}$ denote the predictions for a real neighbor and a pseudo neighbor, 
while $a_{ru_i}$ and $ a_{pu_j}$ represent their corresponding attention scores.
Here, we introduce three approaches to compute these  scores for $a_{ru_i}$, and the same process applies to $a_{pu_j}$: \textbf{\textit{(i)}} average pooling, \textbf{\textit{(ii)}} behavior similarity, and \textbf{\textit{(iii)}} parametric modeling.
Each method is detailed below.
\small{
\begin{equation}
\begin{aligned}    
a_{ru_i} = \begin{cases} \frac{1}{K}, & \text {if average pooling, } \\
\frac{\exp(-\phi(\mathbf{r}_u, \mathbf{r}_{ru_i}))}{\sum_{ru_i \in \mathcal{U}^u_{ru}}\exp(-\phi(\mathbf{r}_u, \mathbf{r}_{ru_i}))}, & \text {if behavior similarity, } \\ 

\frac{\exp((\mathbf{W}_q^T \hat{\mathbf{r}}_u)^T (\mathbf{W}_k^T \hat{\mathbf{r}}_{ru_i}))}{\sum_{ru_i \in \mathcal{U}^u_{ru}}\exp((\mathbf{W}_q^T \hat{\mathbf{r}}_u)^T (\mathbf{W}_k^T \hat{\mathbf{r}}_{ru_i}))}, & \text {if parametric modeling, }\end{cases}
\end{aligned}
\label{eq:att}
\end{equation}
}\normalsize
where $\mathbf{W}_q, \mathbf{W}_k \in \mathbb{R}^{|\mathcal{I}| \times d}$ are parametric weights, with $d$ as the dimensionality. 
We elaborate on the advantages of each approach: \textbf{\textit{(i)}} Average pooling offers straightforward implementation due to its simplicity. 
\textbf{\textit{(ii)}} Pre-calculated distances based on behavior similarity from Eq.~\eqref{make_user} enhance reusability and better reflect actual user similarity by leveraging the observed interaction history ($\mathbf{r}_{u}$).
\textbf{\textit{(iii)}} Parametric modeling effectively learns user relationships by using cascaded reconstructed interactions ($\hat{\mathbf{r}}_{u}$).
In summary, our method offers flexible selection with a focus either on simplicity, efficiency, or effectiveness, providing DR with complementary knowledge to reconstruct nuanced user preferences.


\vspace{0.1cm}
\smallsection{Objective Function.}
Using the fine-grained representation of user preferences ${\hat{\mathbf{r}}_u}'$ derived in Eq. (\ref{eq:fg_obj}), the final objective of \proposed is reformulated from Eq.~\eqref{eq:diffrec_loss} as follows:
\begin{equation}    
\begin{aligned}
    \mathcal{L}_t = &\,\mathbb{E}_{\boldsymbol{q}\left(\textbf{x}_t \mid \textbf{x}_0\right)}\left[C\left\|{\hat{\mathbf{r}}_u}'-\textbf{r}_u\right\|_2^2\right].
\end{aligned}
\end{equation}

%% file: Table/dataset.tex

\begin{table}[t]
\renewcommand{\arraystretch}{0.9} 
\caption{Statistics of the three public datasets.}
\vspace{-1.0em}
\resizebox{0.48\textwidth}{!}{
\begin{tabular}{ccccc}
\toprule
\textbf{Dataset}                        & \textbf{\#Users} & \textbf{\#Items} & \textbf{\#Interactions} & \textbf{Sparsity (\%)} \\ \hline\hline
\textbf{Yelp }        & 26,695  & 20,220  & 942,328        & 99.83 \\
\textbf{AM-Game } & 2,343   & 1,700   & 39,263         & 99.01 \\
\textbf{Citeulike-t}  & 7,947   & 25,975  & 132,275        & 99.94 \\ \bottomrule
\end{tabular}
}
\vspace{-0.5cm}
\label{Dataset}
\end{table}

%% file: Section/Experiment.tex
\subsection{Experimental Settings}
\subsubsection{\textbf{Datasets.}}
We use three public real-world datasets: Yelp\footnote{\url{https://www.yelp.com/dataset}}, Amazon-Game\footnote{\url{https://cseweb.ucsd.edu/~jmcauley/datasets/amazon_v2}}, and Citeulike-t\footnote{\url{https://github.com/js05212/citeulike-t}}. For implicit feedback, we binarize the rating scores of four and five. We then randomly divide the interactions of each user into train/validation/test sets with an 80\% /10\%/10\% split. Table~\ref{Dataset} summarizes the detailed dataset statistics.
\vspace{-0.1cm}
\subsubsection{\textbf{Baselines.}}
We compare \proposed with various baselines: traditional CF methods (e.g., BPRMF~\cite{rendle2009bpr} and LightGCN~\cite{he2020lightgcn}), generative model-based CF methods (e.g., MultiVAE~\cite{liang2018variational} and DiffRec~\cite{wang2023diffusion}), a generative method using neighborhood information (e.g., Ease \cite{steck2019embarrassingly}), and a generative method that incorporates item content features (e.g., ConVAE~\cite{carraro2022conditioned}). 
For ablation analysis, we evaluate three versions of \proposed by integrating \textbf{\textit{(i)}} only real neighbors, \textbf{\textit{(ii)}} only pseudo neighbors, \textbf{\textit{(iii)}} and both, as shown in Table~\ref{tab:main_table}.


\vspace{-0.1cm}
\subsubsection{\textbf{Implementation Details.}}
We conduct a grid search on the validation set utilizing PyTorch with CUDA from RTX A6000 and AMD EPYC 7313 CPU.
Our backbone is an MLP with a hidden size of 1,000.
$\alpha, \beta, \text{ and } \gamma$ are explored in the ranges $\{0.1, 0.3, 0.5, 0.7, 0.9\}$. 
The number of top-$K$ neighbors is searched within $\{10, 20, 50\}$. 
The learning rate and weight decay are searched within $[5\mathrm{e}{-5}, 5\mathrm{e}{-4}]$ and $[0.0, 1\mathrm{e}{-4}]$, respectively.
For evaluation, we use Recall@20 (R@20) and NDCG@20 (N@20).
Note that we report results using attention scores based on behavior similarity, which outperforms other approaches (Eq.~\eqref{eq:att}).
The specific parameters of DR are searched 
following the settings in ~\cite{wang2023diffusion}.
All remaining hyperparameters of \proposed are set to default values.
The baseline-specific hyperparameters adhere to the search ranges of original papers.


\input{Table/small_main_table}

\subsection{\textbf{Performance Comparison}}
Table \ref{tab:main_table} shows that \proposed outperforms all baselines across three datasets.
The results are analyzed from three perspectives: 
\textbf{\textit{(i)}} Generative methods generally surpass traditional ones (i.e., BPRMF and LightGCN), except on Citeulike-t, by effectively capturing the complex and diverse interests of users.
\textbf{\textit{(ii)}} ConVAE, which integrates item features into MultiVAE, outperforms MultiVAE, emphasizing the importance of leveraging item-side information.
\textbf{\textit{(iii)}}  Ease, which incorporates neighborhood information into the generative model,  achieves the best performance among competitors, except on AM-game, highlighting the value of neighborhood-based signals.

Building upon the three aforementioned strengths, \proposed outperforms competitors by effectively mitigating personalized information loss through the integration of item content and collaborative signals from personalized neighbors. 
Notably, CDiff4Rec exhibits a significant performance improvement over DiffRec when employing real neighbors, pseudo neighbors, or both.
This indicates that CDiff4Rec provides DR with complementary knowledge, thereby effectively reconstructing nuanced user preferences.


\input{Table/acc_effi}
\input{Table/ablation_topk}

\subsection{\textbf{Study of \proposed}}
\subsubsection{\textbf{Accuracy and efficiency analysis.}}
We compare the accuracy and efficiency of DiffRec and \proposed using Recall (R), NDCG (N) at 10, 50, 100.
Here, wall time is the total elapsed time for model execution, including both training and inference. 
Table~\ref{tab:acc_eff} shows that \proposed consistently outperforms DiffRec across the three datasets with negligible additional overhead.
An analysis of each dataset is as follows:
On Yelp, \proposed improves R@10 by 12.3\% and N@10 by 12.1\%, while wall time increases by just 10.5\%, showing a favorable trade-off between accuracy and computational cost.
On Amazon-Game (AM-Game), \proposed consistently achieves higher scores across all metrics, despite having a wall time similar to that of DiffRec.
On Citeulike-t, \proposed maintains comparable accuracy yet converges faster, indicating improved efficiency.
Overall, these results suggest that \proposed achieves a better balance between accuracy and efficiency than DiffRec.

\subsubsection{\textbf{Hyperparameter study.}}
We investigate the impact of varying  the number of \textbf{\textit{(i)}} top-$K$ neighbors and \textbf{\textit{(ii)}} pseudo-users (i.e., review words).
Here, pseudo-users are selected based on the highest average TF-IDF scores across multiple items, computed as follows:
\vspace{-0.1cm}
\begin{equation}
    \begin{aligned}
        \text{avg}_\text{TF-IDF}(f) = \frac{1}{|\mathcal{I}|}\Sigma^\mathcal{|I|}_{i=1}\text{TF-IDF}(f,i).
    \end{aligned}
\end{equation}
Table \ref{topk} presents that \proposed overall shows superior performance with top-$20$ neighbors and 1,000 pseudo-users across the three datasets.
This result suggests that CDiff4Rec effectively balances accuracy and efficiency, achieving strong performance even with a relatively small number of top-$K$ neighbors and pseudo-users.


%% file: Table/small_main_table.tex
\begin{table}[t!]
    \renewcommand{\arraystretch}{0.9} 

    \caption{The overall performance comparison.  *, **, and *** denote \(p \leq 0.05\), \(p \leq 0.005\), and \(p \leq 0.0005\) for the paired t-test on \proposed with the best baseline, respectively.}
    \captionsetup[subtable]{position=top}
    \vspace{-1em}
    \resizebox{0.4\textwidth}{!}{ 
    \begin{subtable}[t]{0.45\textwidth} 
    \centering
        \begin{tabular}{ccc}
            \toprule
\textbf{Yelp}                        & \textbf{R@20 }                 & \textbf{N@20 }                 \\ \midrule\midrule
LightGCN                             & 0.1047                & 0.0553                \\
BPRMF                                   & 0.0918                & 0.0485                \\
Ease                                 & {\ul 0.1099} & {\ul 0.0601} \\
MultiVAE                             & 0.1056                & 0.0548                \\
ConVAE                               & 0.1082                & 0.0564                \\
\rowcolor[HTML]{FFFFFF} 
DiffRec                     & 0.1045       & 0.0563       \\ \hline
\textbf{Ours (+Real users)}           & 0.1090                 & 0.0590                 \\
\textbf{Ours (+Pseudo-users)}         & 0.1074                & 0.0584                \\
\textbf{Ours (+Real \& Pseudo users)} & \textbf{0.1145***}    & \textbf{0.0622*}      \\ \bottomrule
\end{tabular}
    \end{subtable}
    }
    
    \vspace{0.5em}
    \resizebox{0.415\textwidth}{!}{ 
    \begin{subtable}[t]{0.45\textwidth} 
        \centering
        \begin{tabular}{ccc}
\textbf{Amazon-Game}                 & \textbf{R@20}                                     & \textbf{N@20}                                     \\ \hline\hline
LightGCN                             & \cellcolor[HTML]{FFFFFF}0.1899           & \cellcolor[HTML]{FFFFFF}0.0854           \\
BPRMF                                   & \cellcolor[HTML]{FFFFFF}0.1839           & \cellcolor[HTML]{FFFFFF}0.0844           \\
Ease                                 & \cellcolor[HTML]{FFFFFF}0.2108  & \cellcolor[HTML]{FFFFFF}0.1012  \\
MultiVAE                             & \cellcolor[HTML]{FFFFFF}0.2181           & \cellcolor[HTML]{FFFFFF}0.0995           \\
ConVAE                               & \cellcolor[HTML]{FFFFFF}0.2192           & \cellcolor[HTML]{FFFFFF}0.0997           \\
\rowcolor[HTML]{FFFFFF} 
DiffRec                     & {\ul 0.2193}                    & {\ul 0.1034}                    \\ \hline
\textbf{Ours (+Real users)}           & \cellcolor[HTML]{FFFFFF}0.2204           & \cellcolor[HTML]{FFFFFF}0.1033           \\
\textbf{Ours (+Pseudo-users)}         & \cellcolor[HTML]{FFFFFF}0.2250   & \cellcolor[HTML]{FFFFFF}\textbf{0.1054*} \\
\textbf{Ours (+Real \& Pseudo users)} & \cellcolor[HTML]{FFFFFF}\textbf{0.2255*} & \cellcolor[HTML]{FFFFFF}0.1052  \\ \bottomrule
\end{tabular}
    \end{subtable}
    }

\vspace{0.5em}
\resizebox{0.395\textwidth}{!}{ 
    \begin{subtable}[t]{0.45\textwidth} 
        \centering
        \begin{tabular}{ccc}
\textbf{Citeulike-t}                 & \textbf{R@20}                                      & \textbf{N@20}                                       \\ \hline\hline
LightGCN                             & \cellcolor[HTML]{FFFFFF}0.1574   & \cellcolor[HTML]{FFFFFF}0.0800    \\
BPRMF                                   & \cellcolor[HTML]{FFFFFF}0.1535            & \cellcolor[HTML]{FFFFFF}0.0792             \\
Ease                                 & \cellcolor[HTML]{FFFFFF}{\ul 0.1591}   & \cellcolor[HTML]{FFFFFF}{\ul 0.0852}    \\
MultiVAE                             & \cellcolor[HTML]{FFFFFF}0.1447            & \cellcolor[HTML]{FFFFFF}0.0770             \\
ConVAE                               & \cellcolor[HTML]{FFFFFF}0.1477            & \cellcolor[HTML]{FFFFFF}0.0775             \\
\rowcolor[HTML]{FFFFFF} 
DiffRec                     & 0.1491                     & 0.0791                      \\ \hline
\textbf{Ours (+Real users)}           & \cellcolor[HTML]{FFFFFF}0.1539            & \cellcolor[HTML]{FFFFFF}0.0822             \\
\textbf{Ours (+Pseudo-users)}         & \cellcolor[HTML]{FFFFFF}0.1605   & \cellcolor[HTML]{FFFFFF}0.0864    \\
\textbf{Ours (+Real \& Pseudo users)} & \cellcolor[HTML]{FFFFFF}\textbf{0.1616**} & \cellcolor[HTML]{FFFFFF}\textbf{0.0885***} \\ \bottomrule
\end{tabular}
\end{subtable}
}
\vspace{-0.4cm}
\label{tab:main_table}
\end{table}

%% file: Table/acc_effi.tex
\begin{table}[t]
\renewcommand{\arraystretch}{1.1} 
\setlength{\abovecaptionskip}{1pt}   
\centering
\renewcommand{\tabcolsep}{0.5mm} 
\caption{Acc-efficiency comparison: DiffRec vs. CDiff4Rec.}
\resizebox{\linewidth}{!}{%
\begin{tabular}{c|c|ccccccc}
\toprule
\textbf{Dataset}                       & \textbf{Model}                             & \textbf{R@10}                  & \textbf{R@50}                  & \textbf{R@100}                 & \textbf{N@10}                  & \textbf{N@50}                  & \textbf{N@100}                 & \textbf{\begin{tabular}[c]{@{}c@{}}Wall\\time\end{tabular}}              \\ \hline\hline
                                       & \textbf{DiffRec}                           & 0.0643                         & 0.1882                         & 0.2818                         & 0.0437                         & 0.0779                         & 0.0982                         & 0:37:47                         \\
\multirow{-2}{*}{\textbf{Yelp}}        & \cellcolor[HTML]{EFEFEF}\textbf{CDiff4Rec} & \cellcolor[HTML]{EFEFEF}0.0722 & \cellcolor[HTML]{EFEFEF}0.2027 & \cellcolor[HTML]{EFEFEF}0.2966 & \cellcolor[HTML]{EFEFEF}0.0490 & \cellcolor[HTML]{EFEFEF}0.0853 & \cellcolor[HTML]{EFEFEF}0.1057 & \cellcolor[HTML]{EFEFEF}0:41:45 \\ \hline
                                       & \textbf{DiffRec}                           & 0.1365                         & 0.3542                         & 0.4689                         & 0.0771                         & 0.1279                         & 0.1478                         & 0:25:13                         \\
\multirow{-2}{*}{\textbf{AM-Game}}     & \cellcolor[HTML]{EFEFEF}\textbf{CDiff4Rec} & \cellcolor[HTML]{EFEFEF}0.1442 & \cellcolor[HTML]{EFEFEF}0.3577 & \cellcolor[HTML]{EFEFEF}0.4813 & \cellcolor[HTML]{EFEFEF}0.0806 & \cellcolor[HTML]{EFEFEF}0.1299 & \cellcolor[HTML]{EFEFEF}0.1514 & \cellcolor[HTML]{EFEFEF}0:25:54 \\ \hline
                                       & \textbf{DiffRec}                           & 0.1175                         & 0.2270                         & 0.2826                         & 0.0748                         & 0.1008                         & 0.1109                         & 3:00:25                         \\
\multirow{-2}{*}{\textbf{Citeulike-t}} & \cellcolor[HTML]{EFEFEF}\textbf{CDiff4Rec} & \cellcolor[HTML]{EFEFEF}0.1177 & \cellcolor[HTML]{EFEFEF}0.2282 & \cellcolor[HTML]{EFEFEF}0.2838 & \cellcolor[HTML]{EFEFEF}0.0760 & \cellcolor[HTML]{EFEFEF}0.1019 & \cellcolor[HTML]{EFEFEF}0.1120 & \cellcolor[HTML]{EFEFEF}2:55:51 \\ \bottomrule
\end{tabular}
}
\vspace{-0.2cm}
\label{tab:acc_eff}
\end{table}

%% file: Table/ablation_topk.tex
\begin{table}[t]
\renewcommand{\arraystretch}{0.85} 
\caption{Impact of varying the number of top-$K$ neighbors (rows) and pseudo-users (columns) across three datasets.}
\vspace{-1.0em}
\resizebox{0.85\linewidth}{!}{%
\begin{tabular}{c|c|cccc}
\toprule
\multirow{2}{*}{\textbf{Dataset}}              & \multirow{2}{*}{\textbf{\#Top-$K$}} & \multicolumn{4}{c}{\textbf{\#Pseudo-users $(|\mathcal{F}|)$}}                                      \\ \cline{3-6} 
                                      &                         & \textbf{1,000}  & \textbf{5,000} & \textbf{10,000} & \textbf{20,000} \\ \hline\hline
\multirow{3}{*}{\textbf{Yelp}}        & \textbf{10}             & 0.1092          & 0.1077         & 0.1091          & 0.1115          \\
                                      & \textbf{20}             & 0.1123          & 0.1120          & 0.1117          & 0.1131          \\
                                      & \textbf{50}             & \textbf{0.1145} & 0.1136         & 0.1132          & 0.1133          \\ \hline
\multirow{3}{*}{\textbf{AM-Game}} & \textbf{10}             & 0.2134          & 0.2128         & 0.2151          & 0.2196          \\
                                      & \textbf{20}             & \textbf{0.2255} & 0.2155         & 0.2147          & 0.2142          \\
                                      & \textbf{50}             & 0.2207          & 0.2169         & 0.2191          & 0.2181          \\ \hline
\multirow{3}{*}{\textbf{Citeulike-t}} & \textbf{10}             & 0.1595          & 0.1594         & 0.1591          & 0.1603          \\
                                      & \textbf{20}             & 0.1584          & 0.1605         & \textbf{0.1616} & 0.1607          \\
                                      & \textbf{50}             & 0.1585          & 0.1566         & 0.1596          & 0.1572          \\ \bottomrule
\end{tabular}
}
\vspace{-0.4cm}
\label{topk}
\end{table}

%% file: Section/Conclusion.tex
In this paper, we propose \proposed, a novel diffusion-based recommender system that integrates item content and collaborative signals to mitigate the loss of personalized information. 
By converting item review words into pseudo-users and incorporating the preference information from both real and pseudo personalized neighbors, \proposed effectively reconstructs nuanced user preferences through complementary knowledge.
Experimental results on three public datasets show its superiority, 
while additional analyses suggest that  \proposed achieves a good balance between accuracy and efficiency.
We expect that \proposed will inspire further research on leveraging diverse item-side information to advance existing user-oriented DR.
